\documentclass[sigconf, manuscript, nonacm]{acmart}
\AtBeginDocument{%
  \providecommand\BibTeX{{%
    \normalfont B\kern-0.5em{\scshape i\kern-0.25em b}\kern-0.8em\TeX}}}

\usepackage{multicol}
\begin{document}

\title{PartiPlay: A Participatory Game Design Kit for Neurodiverse Classrooms}

\author{Patricia Piedade}
\affiliation{%
  \institution{Interactive Technologies Institute, University of Lisbon}
  \city{Lisbon}
  \country{Portugal}}
\email{patricia.piedade@tecnico.ulisboa.pt}

\author{Isabel Neto}
\affiliation{%
  \institution{INESC-ID, University of Lisbon}
  \city{Lisbon}
  \country{Portugal}}
\email{isabel.neto@tecnico.ulisboa.pt}

\author{Ana Pires}
\affiliation{%
  \institution{Interactive Technologies Institute, University of Lisbon}
  \city{Lisbon}
  \country{Portugal}}
\email{ana.pires@iti.larsys.pt}

\author{Rui Prada}
\affiliation{%
  \institution{INESC-ID, University of Lisbon}
  \city{Lisbon}
  \country{Portugal}}
\email{rui.prada@tecnico.ulisboa.pt}

\author{Hugo Nicolau}
\affiliation{%
    \institution{Interactive Technologies Institute, University of Lisbon}
  \city{Lisbon}
  \country{Portugal}}
\email{hugo.nicolau@tecnico.ulisboa.pt}
\renewcommand{\shortauthors}{Piedade, et al.}

\begin{abstract}
  Play is a central aspect of childhood development, with games as a vital tool to promote it. However, neurodivergent children, especially those in neurodiverse environments, are underserved by HCI games research. Most existing work takes on a top-down approach, disregarding neurodivergent interest for the majority of the design process. Co-design is often proposed as a tool to create truly accessible and inclusive gaming experiences. Nevertheless, co-designing with neurodivergent children within neurodiverse groups brings about unique challenges, such as different communication styles, sensory needs and preferences. Building upon recommendations from prior work in neurodivergent, mixed-ability, and child-led co-design, we propose a concrete participatory game design kit for neurodiverse classrooms: PartiPlay. Moreover, we present preliminary findings from an in-the-wild experiment with the said kit, showcasing its ability to create an inclusive co-design process for neurodiverse groups of children. We aim to provide actionable steps for future participatory design research with neurodiverse children.
\end{abstract}




\begin{teaserfigure}
  \includegraphics[width=\textwidth]{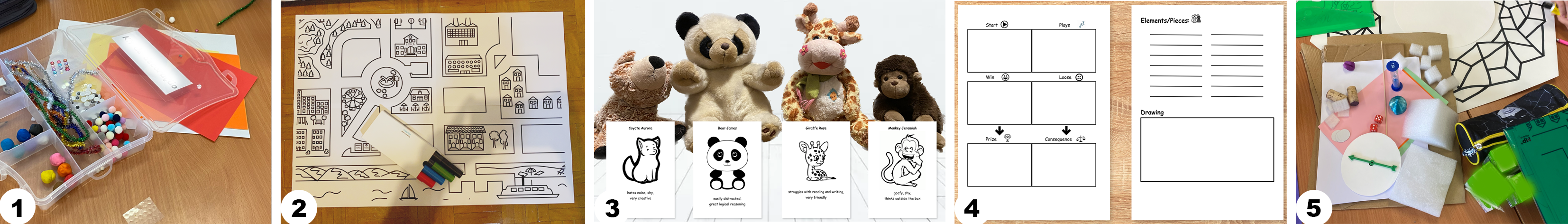}
  \caption{Participatory game design kit. 1-5 depict sessions 1-5.}
  \Description{The figure is composed of 5 images. From left to right, 1 depicts the robot decorating kit from session 1, a clear plastic box with dividers which contain plasticine, pipe cleaners, and stickers. 2 depicts the map for the storytelling activity of session 2, a A2 size piece of paper with a two-dimentional map printed on it and a box of markers on top. 3 depicts the proxies from session 3, four stuffed animals (a coyote, a panda, a giraffe and a monkey) with their character sheets in front of them (pieces of paper with a cartoon animal, its name and nuerodivergent traits). 4 depicts the worksheets from session 4. 5 depicts the prototyping kit from session 5.}
  \label{fig:teaser}
\end{teaserfigure}


\maketitle

\section{Introduction}
% Importance of Play
Play is a fundamental childhood activity \cite{hirsh-pasek,liu_hirsh_neuroscience}, promoting intellectual, creative and social development \cite{Fromberg1990, Fromberg1992, garvey1990, huizinga2014homo}. Through play, children form friendships and find a space for self-expression and exploration \cite{Fromberg2012, peds, Johnson1987}.
Games are widely used to unlock the benefits of play, offering pleasurable engagement and positive outcomes for players' well-being \cite{Iacovides2019, Jones2014}. Moreover, they have the potential to promote inclusive and equally engaging experiences for players with and without disabilities \cite{inspo}.

% Neurodivergent meaning
From the standpoint of \textit{neurodiversity}, we recognize a multitude of neurological differences in human brains \cite{shakespeare1996, putnam2005}, where most brains are \textit{neurotypical}. Some diverge from these norms, thus, referred to as \textit{neurodivergent} (e.g., ADHD, autism, dyslexia, and intellectual disabilities) \cite{Dalton2013}. 
% Need for neurodivergent and neurodiverse games
As with many marginalized groups within educational settings, access to inclusive play for neurodivergent students remains a challenge \cite{Fromberg2012, ng4}. Furthermore, HCI games research focused on this population primarily creates games for medical and training purposes (i.e., serious games) \cite{prop3}. The main goal of these games is to dress up boring and repetitive activities,  which tend to prioritize training over play, and are intended to be used by neurodivergent players alone, reducing opportunities for neurodiverse play (i.e., play that involves neurodivergent and neurotypical players) and disregarding neurodivergent interests \cite{prop3}.

\begin{figure*}[t]
    \centering
    \includegraphics[width=\textwidth]{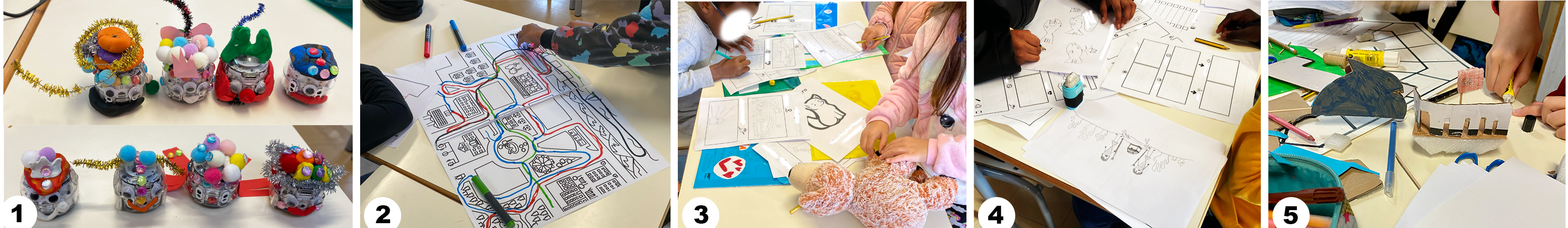}
    \caption{Neurodiverse classroom evaluation of the participatory game design kit. 1-5 depict sessions 1-5.}
    \Description{The figure is composed of 5 images. From left to right, 1 depicts the decorated robots from session 1. 2 depicts the map from session 2 with paths drawn and a robot on it. 3 depicts a group sitting at a table with their stuffed animal proxy and session 3 worksheets. 4 depicts a group sitting at a table with their session 4 worksheets and additional drawings on scratch paper. 5 depicts a child putting together a cardboard boat, while another holds a cardboard shark figure, as part of session 5 low-fidelity prototypes.}
    \label{fig:fig2}
\end{figure*}

% Co-Design as an approach
This paper investigates how to facilitate inclusive play experiences for neurodiverse children in classrooms, leveraging a co-design process as a playful and inclusive experience \cite{nci3}.  
% Eperimental context + Robots
Given the lack of games designed for a neurodiverse context and the potential of robots as game elements within this context \cite{inspo}, we propose a participatory game design kit aimed at neurodiverse classrooms. This toolkit aims to include neurodivergent interests, keeping all children engaged and adapting activities to the different communication styles and needs. At the same time, scaffolding children to develop their social skills, as they needed to interact with each other, discussing and negotiating possible solutions, and converging in a unique group solution in a playful and creative activity.  
Inspired by previous research \cite{inspo}, we explore the potential of the game design kit with 81 neurodiverse children (age 6-12, 19 neurodivergent) from a mainstream school in a four-month process. Children used small robotic devices and the proposed game design kit to create inclusive games through a five-session design process (Figure \ref{fig:fig2}).
% Contributions 
We contribute an inclusive participatory game design kit for neurodiverse groups of children and its evaluation with 81 children within four classrooms. We highlight the effectiveness of methodologies, such as Expanded Proxy Design \cite{prop1} and the importance of physical ownership of design elements.

\section{Game Design Kit}
The methodology described herein considers the scenario of designing a game that utilises an off-the-shelf robot - the Ozobot. Nevertheless, we believe this kit can be employed in projects that do not leverage robots as part of their design process.
Within the context of the neurodiverse classroom, under the advice of educators who operate within it, we employed both small group (4 to 6 children) and individual activities. Our process, based around the Diversity for Design Framework \cite{ng1} and Metatla et al.'s workshop organisation \cite{inspo}, is divided into five hour-and-half sessions, each building upon the last, but with its own goals. 
To mitigate information loss in between sessions, we employ two ongoing methods. The first, \textit{participatory recap}, consists of a researcher prompting the children to retell the happenings of the previous sessions at the beginning of each session. The second consists of children keeping a \textit{project portfolio} (directly inspired by Malinverni et al.'s use of project boxes in a participatory design project with autistic children \cite{mali}) where they store worksheets and artefacts from previous sessions and to which they can refer back at any time.
Throughout the process, we employ several worksheets available as supplementary material (Figure \ref{fig:teaser}). All worksheets include pictograms, text, and enough space to write or draw answers, supporting children who may struggle with reading and writing \cite{Guha2008, ng1}.

\textbf{Session 1: Building Rapport.}
The main goal of session one is to build relationships. We start with an introduction of the research team and project. Following previous recommendations towards making social mechanics explicit \cite{ng4}, we follow this introduction with an icebreaker, using a foam ball to make turn-taking explicit. The ball is passed around the classroom, and whoever holds it shares with the group their name, age, current mood, and one fun fact about themselves. 
%introduction
%icebreaker
To build team spirit within the small groups, which should remain the same throughout the design process, we recommend having them pick out a team name.
Besides building a sense of continuity, Malinverni et al. \cite{mali} highlight the personalised nature of each project box. We propose using a folder rather than a box for ease of storage. To get children acquainted with their portfolio, we propose having them customise it with crafting materials (e.g., plastic A4 folder, blank paper, colouring material and laminating plastic).
%portfolio decoration
Afterwards, to build excitement and familiarity with the technology to be used, in this case, the Ozobot robot, we suggest having the group customise it in a joint crafting activity, which has proven effective in mixed-ability settings \cite{inspo}. Each group should receive a similar decoration kit. We suggest including plasticine, stickers, googly eyes, coloured paper and pipe cleaners (Figure \ref{fig:teaser}.1). If this customised artefact cannot remain with the children after the session, we recommend photographing it for posterity.
%ozobot decoration
Finally, as teachers highlighted that children enjoyed showcasing their work to others, each group would be given the opportunity to present their customised robot to the class. We propose that these presentations remain voluntary, not pressuring any child into public speaking.
%presentation

\textbf{Session 2: Exploring the Technology.}
Session two aims to explore the technological element. Through a series of game-like activities, which tend to be engaging for neurodivergent children in co-design settings \cite{mali}, groups will be able to explore the robot's features.
The first activity explores the Ozobot's ability to follow lines drawn in maker through story-telling. Each group receives an A2 map of a town, where they can draw the Ozobot's path throughout its daily routine. We suggest adding elements to the town that match children's specific interests (e.g., favourite restaurant or sports stadium) per previous guidelines for neurodivergent co-design \cite{ng1} (Figure \ref{fig:teaser}.2). The second utilises a puzzle set containing the same lines the Ozobot can follow. In this problem-solving exercise, children will be asked to build a path with puzzle pieces, from a start piece to a 3D house structure. The last activity also plays into children's interests, which has been shown to promote motivation in neurodivergent children \cite{ng1} by asking each group to pick a song for the Ozobot to dance to on sceneries decorated by each child. 
%introduction
%activity 1
%activity 2
%activity 3

\textbf{Session 3: Expanded Proxy Design for Narrative Building.}
%introduction
The third session shifts the focus from the robot to game design. We propose starting with an icebreaker akin to session one's, having each child share their favourite game. Then, once again, relating to the group through interests, we suggest using some of these games to exemplify the game elements groups will define during this session.
This session takes on Expanded Proxy Design \cite{prop1}, introducing each group to a stuffed animal proxy with neurodivergent characteristics (Figure \ref{fig:teaser}.3). The original methodology suggests the proxy should have embodied characteristics that represent the given minority \cite{prop1}, however, as neurodivergence is most often invisible, we propose using a variety of stuffed animals, disregarding physical attributes. We suggest conveying the neurodivergent characteristics through a worksheet and verbally introducing the stuffed animal to the group. Alongside the proxy stuffed animal, each group should receive a worksheet specifying which game elements to choose. We propose basic narrative elements for this session: setting, goal, obstacles and aids. We add an extra category for the Ozobot's function. Alongside the worksheets, we suggest providing each group with blank paper, promoting a balance between structure and open-ended activities which is beneficial when working with neurodiverse groups \cite{nci3}. We propose that each group present their game and proxy ``friend", for the reasons mentioned in session one and to allow the research team to better understand their game concept.
%expanded proxy
%worksheet
%presentation

\textbf{Session 4: Game Mechanics.}
Session four continues the work of session three, asking the children to further their game concepts with game mechanics. We advise that researchers make clear the difference between narrative game progression and game mechanics emphasising they should now be describing how the game is played by the player rather than a player character.
The worksheet for this session contains the same sections as session three, encouraging children to rethink their narrative elements as mechanics, and additional ones related to mechanic progression: turns, actions, winning, losing, rewards, and consequences (Figure \ref{fig:teaser}.4). The last page of the worksheet allows groups to create a checklist of game pieces to be created in the next session. Once again, extra scratch paper should be made available and can be used for some initial prototyping to aid in the joint decision-making exercise. Similar to session three, we propose that each group present their work to the class at the end of the session.
%introduction
%worksheet
%presentation

\textbf{Session 5: Prototyping and Play-testing.}
The final session closes the design cycle with low-fidelity prototyping and play-testing of the developed game concepts. When introducing the activity, it is important to note which materials children can use.
Each group should be provided with a similar prototyping kit. We suggest including foam, paper, cardboard, dice, pinwheels and hourglasses (Figure \ref{fig:teaser}.5). We recommend analysing children's game piece list from the previous session and providing groups with any materials they specifically request. Each group will have a set time to create their game prototype, then elect a game presenter who will remain at their workstation while classmates come by to play-test their concept. We recommend asking for the aid of the teacher when coordinating the switching of tables for the play-test. Observing or recording the play-tests will provide valuable insights into the game concepts.

%introduction
%prototyping
%play-testing

\section{Classroom Evalutation}

To evaluate this toolkit, we conducted the five proposed co-design sessions with four neurodiverse classrooms. In total, 81 neurodiverse students (aged 6-12), 19 neurodivergent \footnote{13 learning differences, 1 dyslexia, 2 intellectual disabilities, 2 ADHD, 1 Down's Syndrome, and 1 Global Developmental Delay}, participated in co-designing a robotically-enhanced board game. We followed the kit's methodology, making adjustments during and in-between sessions to overcome practical challenges, as suggested by prior work \cite{mali}.
During this evaluation period, researchers wrote daily field notes. Afterwards, we conducted a thematic analysis \cite{braun2012thematic} of said field notes. Two researchers inductively coded the notes. This coding was later peer-validated with others present at the co-design sessions and the project's advisors. We present a list of preliminary findings regarding the practical application of this toolkit in neurodiverse classrooms:

\textbf{Crafting activities promote individual ownership over artefacts and reduce conflict.}
Crafting activities were overall better received by the children when compared to more abstract thought exercises, such as deciding on game elements. Neurodivergent children, in particular, tended to disengage from the latter due to the activities being less appealing to them. Low-fidelity prototyping worked remarkably well. Having decided in session four what game elements they would need to prototype, each child in a group was able to take ownership of a number of game pieces and individually work on them. While neurotypical group members focused on completing all game pieces in the allotted time, neurodivergent children often focused on a single game piece, perfecting it to an impressive degree and receiving praise from group mates (Figure \ref{fig:fig2}.5).

\textbf{Allowing for multiple ways of expression makes for a more equitable experience.}
Previous literature on co-designing with neurodivergent children highlights the need for reading and writing supports \cite{ng1}, this was reinforced by the participating teachers. When first presented with the worksheets, most children opted to write down their answers, as was customary in the classroom. However, upon realizing they were allowed to draw, many switched to this modality (Figure \ref{fig:fig2}.4). In one case, a child with dyslexia became distraught for not being able to finish writing her answers in time but was instantly relieved when realizing she could draw instead. Furthermore, the pictograms, especially those that reoccurred across worksheets, gave neurodivergent children who struggled with reading a means to interpret the tasks independently.

\textbf{Expanded Proxy Design creates game concepts that incorporate neurodivergent traits.}
When speaking with the teachers about our project, most stated that neurodivergent children were not visibly different in their classmates' eyes. The cases of exclusion we observed in the classroom were justified by teachers as neurotypical peers interpreting neurodivergent characteristics as undesirable personality traits. Therefore, Expanded Proxy Design \cite{prop1} was crucial to creating games that centred neurodivergent interests without calling undue attention to neurodivergent children (Figure \ref{fig:fig2}.3). Neurotypical group members remembered their proxy's needs and desires throughout the design process, making explicit remarks about them until the very last session. Neurodivergent children related to the proxies and became advocates for their well-being. 

\textbf{Ensuring children's physical ownership over all design artefacts is essential for true co-design.}
This kit attempts to provide child co-designers with a complete record of the design process through the project portfolio. However, during the sessions, we realized the importance of allowing children to keep the design artefacts they created. Due to the limited amount of Ozobots available, after the first session, we had to remove the children's decorations to allow other classrooms to partake in the activity (Figure \ref{fig:fig2}.1). Upon returning to the classrooms for the following session, children, in particular, neurodivergent children who had spent a lot of time and energy perfecting the Ozobot's decorations, were upset by seeing the robots were bare. We corrected this in the last session by providing Ozobot helmets that children could decorate and keep. At the end of this session, groups laid out their prototypes and enthusiastically decided who would take home each game piece.
\section{Conclusion}

In this work, we take on the lens of neurodiversity, aiming to explore the inclusive potential of co-designing games with neurodivergent and neurotypical children. 
We explore co-design methodologies with this population, creating a participatory game-design kit. We applied the design kit to four different classrooms in a five-session process. We highlighted the importance of adapting the activities, duration, wording, and each deliverable to children's abilities and needs. We gathered several game-design concepts into the co-design process to reach a low-fidelity prototype, play-testing it in neurodiverse classrooms. Our findings suggest that different communication means (e.g., verbal, written, and drawn) allowed children to understand activities, share and discuss their ideas and express their views, reaching a group consensus. Also, using expanded design proxies created an inclusive experience and enriched their empathy toward each other, increasing their openness to different opinions during the processes. The crafting nature of the co-design activities allowed them to develop a sense of ownership that was key for their engagement and playfulness and reduced interpersonal friction and conflict. Furthermore, the physicality of the deliverables allowed them to share the insights from their work with their teammates, their classroom, and even their families. 

Our work builds on previous efforts toward creating inclusive play experiences using a co-design activity. The process offers a new perspective to foster inclusion through play by combining neurodiverse players, co-design, and a classroom group activity. We hope to encourage HCI researchers to explore the inclusive potential of co-designed games for neurodiverse stakeholders.

\begin{acks}

This work was supported by the European project DCitizens: Fostering Digital Civics Research and Innovation in Lisbon (GA 101079116), by the Portuguese Recovery and Resilience Program (PRR), IAPMEI/ANI/FCT under Agenda C645022399-00000057 (eGamesLab) and the Foundation for Science and Technology (FCT) funds
 \newline SFRH/BD/06452/2021 and UIDB/50021/2020.

\end{acks}

\bibliographystyle{ACM-Reference-Format}
\bibliography{biblio}

\appendix

\end{document}